\documentclass[twocolumn,english,aps,graphicx,amsmath,showpacs]{revtex4}
\usepackage[T1]{fontenc}
\usepackage[latin1]{inputenc}
\usepackage{babel}
\usepackage{graphics}

\begin{document}

\title{All-optical production of $^7$Li Bose-Einstein condensation using Feshbach resonances.}
\author{Noam Gross and Lev Khaykovich}
\affiliation{Department of Physics, Bar-Ilan University,
Ramat-Gan, 52900 Israel.}

\begin{abstract}
We show an all-optical method of making $^7$Li condensate using
tunability of the scattering length in the proximity of a Feshbach
resonance. We report the observation of two new Feshbach resonances
on $|F=1,m_{F}=0\rangle$ state. The narrow (broad) resonance of
$7$~G ($34$~G) width is detected at $831\pm4$~G
($884^{+4}_{-13}$~G). Position of the scattering length zero
crossing between the resonances is found at $836\pm4$~G. The broad
resonance is shown to be favorable for run away evaporation which we
perform in a crossed-beam optical dipole trap. Starting directly
form the phase space density of a magneto-optical trap we observe a
Bose-Einstein condensation threshold in less than 3 s of forced
evaporation.
\end{abstract}


\maketitle


\section{Introduction}

Achieving quantum degeneracy in ultracold atomic gases by
all-optical means becomes a well accepted experimental technique
because of several inherent advantages
\cite{Chapman01,Yb_BEC,Cs_BEC,Granade02,Kinoshita05,Dumke06,Fuchs07}.
Optical traps allow strong confinement resulting in high collision
rates and rapid evaporative cooling. Confinement of arbitrary spin
states and spin state mixtures are readily obtained. The possibility
to tune interactions via Feshbach resonances usually requires
optical trapping as they frequently occur in states that cannot be
trapped magnetically \cite{Feshbach_review1,Feshbach_review2}.
Finally, large current coils needed for magnetic-field trapping that
restrict optical access are avoided.

The first successful demonstration of an all-optically achieved
$^{87}$Rb Bose-Einstein condensation (BEC) allowed significant
increase in the rate of BEC production and the resulting condensates
were F=1 spinors \cite{Chapman01}. However, the main driving force
behind the search for all-optical techniques was the need to
condense specific atoms where the 'conventional' evaporation in the
magnetic trap was not possible. Two prominent examples are spinless,
and thus magnetically untrappable, BEC of Yb atoms achieved in a
doubled YAG crossed dipole trap \cite{Yb_BEC} and a BEC of
$^{133}$Cs atoms \cite{Cs_BEC} for which the strongly enhanced
two-body losses from the magnetically trappable states prevent the
condensate formation in the 'standard' way \cite{Dalibard98}.

Although $^7$Li atoms can be evaporatively cooled in a magnetic trap
\cite{Hulet95}, the task remains challenging due to several reasons.
First, $^7$Li atoms posses a relatively small scattering length
($a=-27a_{0}$, where $a_{0}$ is the Bohr radius) and high two-body
loss rate \cite{Li_a}. Second, the initial phase space density is
unfavorably limited by the absence of polarization-gradient cooling
mechanism. Third, since the scattering length drops with increased
temperature and crosses zero at $T=8$ mK \cite{zero_a}, the use of
adiabatic compression to increase the elastic collisional rate is
ineffective. Therefore, the strong magnetic confinement needed to
keep evaporation time comparable with heavier alkalies, such as Rb
and Na, requires the design of a miniaturized trap. This is done by
either a small-volume vacuum chamber and high currents
\cite{Schreck01} or a vacuum compatible minitrap \cite{Wang07} which
both increase the experimental complexity. Finally, even if the
strong confinement is achieved the scattering length is still
negative which prevents the formation of a stable BEC.

In this paper we show an all-optical way to condense $^7$Li atoms
using tunable interatomic interactions. We use a 100 W Ytterbium
fiber laser to produce $\sim$2 mK potential well which traps
$\sim$10$^{6}$ atoms from a magneto-optical trap (MOT). We explore
two new Feshabach resonances on $|F=1, m_F=0\rangle$ state and find
that one of them is favorable for the efficient forced evaporation
starting directly from the phase space density achieved in the MOT.
We obtain a BEC on $|F=1, m_F=0\rangle$ internal state in less than
3 s of evaporation time.


\section{Experimental details}

The $^7$Li atoms' route to quantum degeneracy starts in an oven
where they are heated up to $450^{0}$C to increase their vapor
pressure. An atomic beam is collimated by two distant apertures and
slowed down in an increased magnetic-field Zeeman-slower. The
capture velocity of the slower is set to $\sim$800 m/s.

Our MOT design and parameters are similar to that described in ref
\cite{MOT98}. After a loading time of 3 s, the trap contains
$\sim10^{9}$ atoms at a temperature of $1.2$ mK. Trap life-time,
limited by vacuum, is measured to be 10 s. The MOT parameters, such
as pump and repump detunings and magnetic field gradient, are
optimized to maximize the number of atoms. To improve the initial
phase space density we implement a so called compressed MOT (CMOT)
stage \cite{MOT98,Mewes99}. For $50$ ms the laser intensities are
attenuated, detunings are decreased and the magnetic field gradient
is increased. As a result the temperature is reduced to 300 $\mu$K
and nearly half of the atoms are lost. By the end of this phase
$\sim5\times10^{8}$ atoms are left in the trap with density of
$n=4\times10^{11}$ atoms/cm$^{3}$ and phase-space density of $\rho
=2\times10^{-5}$.

\begin{figure}

{\centering \resizebox*{0.47\textwidth}{0.23\textheight}
{{\includegraphics{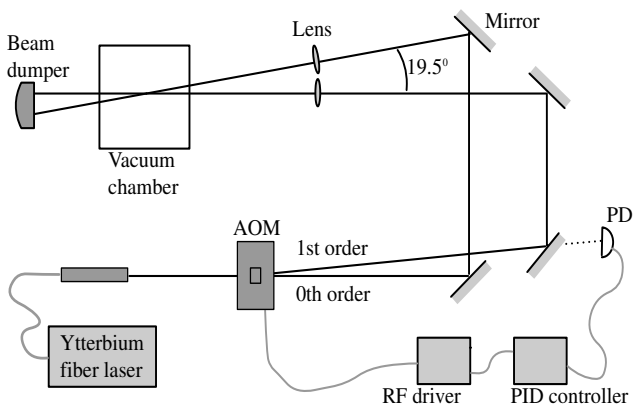}}}
\par}
\caption{\label{apparatus}A schematic representation of the
optical-dipole trap experimental setup. The first order diffraction
beam is focused to a waist of 31 $\mu$m and aligned to the MOT
center. The zeroth order beam is weakly focused to 300 $\mu$m. It
crosses the first order beam at an angle of 19.5$^{0}$ and it only
becomes important at the final stages of evaporation. AOM -
Acousto-optic modulator; PD - Photodetector. }
\end{figure}

The realization of the optical-dipole trap is shown in Fig.
\ref{apparatus}. A CW Ytterbium fiber laser generates $100$ W of
linearly polarized light at 1.07 $\mu$m. The first order diffraction
beam of an acousto-optic modulator (AOM), with $80 \%$ diffraction
efficiency, is focused at the center of the MOT to a waist of 31
$\mu m$. The zero order diffraction beam is only weakly focused to
300 $\mu$m and it crosses the first order beam at an angle of
19.5$^0$. When maximum RF power is applied to the AOM this beam is
very weak and it has no effect on the atoms. At full laser power (65
W in the trap region) the trap depth, produced by the first order
diffraction beam, is estimated to be $\sim2$ mK. Oscillation
frequencies in the trap are predicted to be
$\omega_{r}=2\pi\times16$ kHz and $\omega_{z}=2\pi\times120$ Hz for
the radial and longitudinal directions, respectively. The radial
oscillation frequency was measured by parametric driving technique
and found to be in excellent agreement with the predicted one. We
keep the first order diffraction beam at its full power throughout
the MOT loading time and atoms are loaded into the dipole trap
mainly during the CMOT phase. Mechanical shutters are then used to
block all resonant light from reaching the trapped atoms. The MOT
pump beam, associated with the $|F=2\rangle$ ground state, is
blocked only a few milliseconds later to allow a $1$ ms optical
pumping pulse, effectively transferring all atoms into the
$|F=1\rangle$ ground state. Eventually, we are left with
$\sim10^{6}$ trapped atoms. Phase-space density conditions are
similar to those resulted from the CMOT phase.

We note that the loading efficiency of the optical trap from the
CMOT is rather poor due to the small spatial overlap between the two
traps. Even for 100 W laser power the high initial temperature of
the $^7$Li atoms oblige tight focusing of the beam in order to
create a deep enough potential. This makes the use of a compressible
crossed dipole trap to improve the loading conditions impractical
\cite{Kinoshita05}.


\begin{figure}

{\centering \resizebox*{0.47\textwidth}{0.27\textheight}
{{\includegraphics{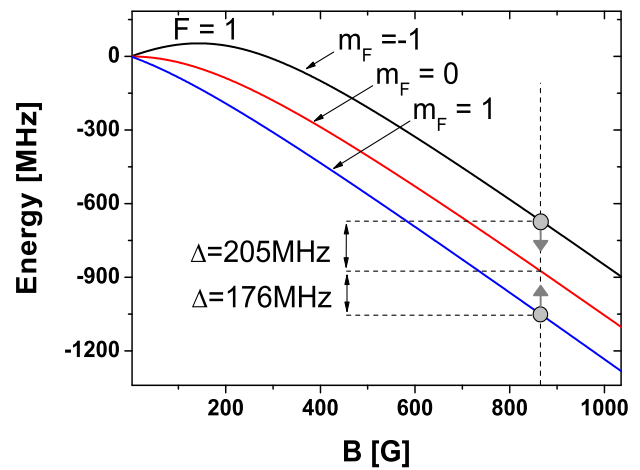}}}
\par}
\caption{\label{Li7F1}Zeeman splitting of the $|F=1\rangle$
hyper-fine state of $^{7}$Li atoms. With the presence of high
magnetic fields, spin-flip collisions between $|m_{F}=-1\rangle$ and
$|m_{F}=1\rangle$ states result in their transfer into
$|m_{F}=0\rangle$ with an excess of kinetic energy equal to 29 MHz
($\sim1.4$ mK).}
\end{figure}

\section{Detection of Feshbach resonances on $|F=1,m_{F}=0\rangle$ state}

The scattering length of $^7$Li atoms on all sublevels and sublevel
mixtures of the $|F=1\rangle$ ground state at zero magnetic field is
positive but very small ($\sim10a_{0}$), impeding efficient
evaporation cooling. However, each sublevel and their mixtures
posses at least one Feshbach resonance \cite{Servaas}. The resonance
on the absolute ground state $|F=1,m_{F}=1\rangle$ has been
previously used for the final stage of forced evaporation in an
optical trap after precooling in a magnetic trap
\cite{bs_ENS,bs_Rice}. The resonances on other sublevels or their
mixtures were neither reported nor used before.

The search for Feshbach resonances requires high offset magnetic
fields. We use two pairs of Helmholtz coils that allow variable bias
fields of up to $1200$ G. When the magnetic field is ramped up to
high values we observe $\sim50\%$ reduction in atom number. Our
state selective measurement shows that the remaining atoms are all
on $|F=1,m_{F}=0\rangle$ state. We attribute this loss to spin-flip
collisions between atoms on $|F=1,m_{F}=-1\rangle$ and
$|F=1,m_{F}=1\rangle$ states. The energy level splitting diagram of
the $|F=1\rangle$ state is shown in Fig. \ref{Li7F1}. A spin-flip
collision that takes place at high magnetic fields flips the
colliding atoms to $|F=1,m_{F}=0\rangle$ state and leaves them with
an excess of kinetic energy equal to $1.4$ mK, which is comparable
to the trap depth. Most of the atoms escape but we observe some
increase in $|m_{F}=0\rangle$ population. Detection of losses and
spontaneous spin polarization are the first manifestations of the
increase in collisional cross-section. As $|m_{F}=0\rangle$ state is
stable against spin-exchange collisions at high magnetic fields, we
search for signatures of Feshbach resonances on this state.


\begin{figure}[htp]

{\centering \resizebox*{0.47\textwidth}{0.67\textheight}
{{\includegraphics{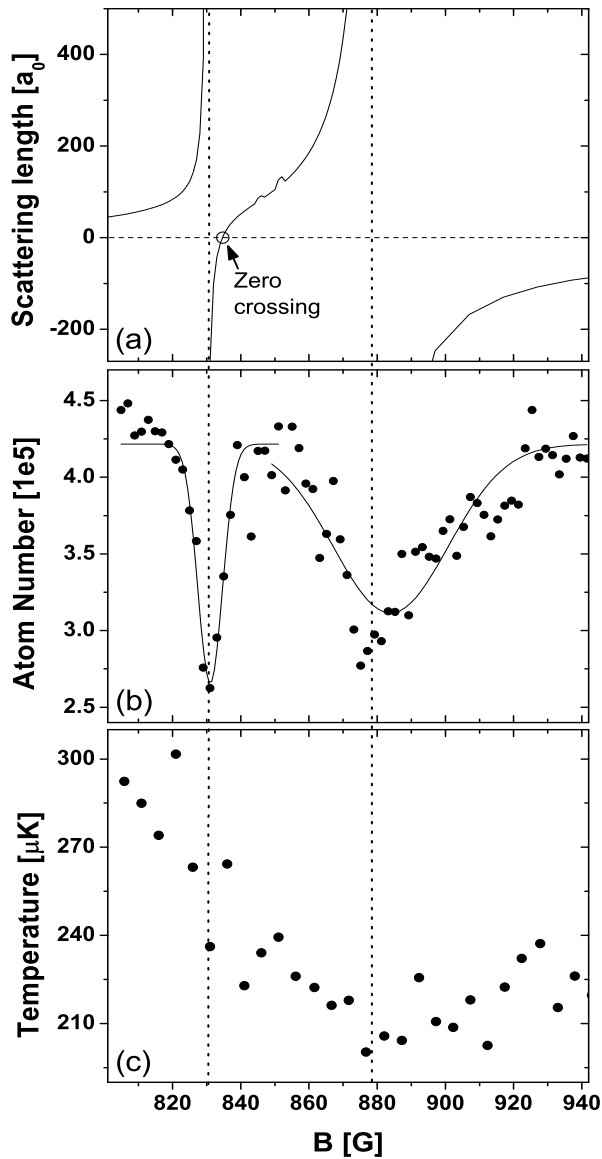}}}
\par}
\caption{\label{feshbach}(a) Theoretical predictions for the
scattering length as a function of bias magnetic field in the
$|F=1;m_{F}=0\rangle$ state \cite{Servaas}. Divergence of the
scattering length signify the presence of two Feshbach resonances.
(b) The resonances are detected by measuring atom loss due to
inelastic collisions. The solid lines are Gaussian fits to define
resonances' positions. Note that the maximal loss does not coincide
with the minimum of the gaussian fit for the broad resonance. (c)
Temperature of the atomic cloud in the optical trap as a function of
bias magnetic field. Decrease in temperature indicates cooling by
free evaporation.}
\end{figure}

Fig. \ref{feshbach}(a) shows a theoretical prediction of the
scattering length as a function of bias magnetic field. Two Feshbach
resonances between $800$ G and $900$ G are indicated by the
divergence of the scattering length. Experimentally, the resonances
are usually observed by detection of atom loss as in their proximity
the inelastic collisional rate is strongly enhanced
\cite{Fedichev96,FR_first_demonstration}. For this measurement we
ramp up the magnetic field to different values in $40$ ms and wait
for $0.5$ s before decreasing it to $386$ G in $40$ ms where
\textit{in situ} state selective absorption imaging is performed. In
Fig. \ref{feshbach}(b) the atom number as a function of bias
magnetic field is shown. The atoms are initially prepared at a
temperature of 300 $\mu$K. Strong losses are observed around $830$ G
and $880$ G featuring a narrow and a broad structures. By fitting
them with simple Gaussian functions, a coarse estimation of the
resonances locations and widths can be determined. According to the
fit, the narrow resonance is $7$ G wide ($1/e^{2}$ radius) and it is
located at $831\pm4$ G. The uncertainty in its position is due to an
uncertainty in the magnetic field calibration which is obtained by
the detection of optical resonances and thus limited by the
linewidth of the excited states and the laser locking quality. The
broad resonance located at $884^{+4}_{-13}$ G is $34$ G wide and
features a notable asymmetric profile which tends to shift the
center of a simple gaussian fit to a higher magnetic field value as
compared to the maximal loss position (detected at 875 G). Such an
asymmetry in losses in the vicinity of broad Feshbach resonances has
been recently reported for $^{39}$K atoms \cite{D'Errico07} and was
attributed to larger three-body loss coefficient from the negative
scattering length side of the resonance and mean field effects. The
followed study of molecule association showed that the position of a
broad Feshbach resonance is indeed shifted to a lower value
\cite{D'Errico07}. Comparison between the experimental measurements
(Fig. \ref{feshbach}(b)) and the theoretical calculations (Fig.
\ref{feshbach}(a)) shows that the maximal loss position suits better
the predicted distance between the two resonances. We therefore
believe that the actual location of the broad resonance is at
somewhat lower value than that given by the Gaussian fit and this
systematic error increases the error bars of the experimentally
detected resonance position.

Most excitingly, we observe an onset of free evaporation cooling
together with the enhanced inelastic losses ( Fig.
\ref{feshbach}(b)). In Fig. \ref{feshbach}(c) we show the
temperature change as a function of the bias magnetic field. A
decrease in the temperature indicates cooling which we attribute to
the establishment of free evaporation when the collision rate
becomes high enough. The observed cooling allows us to decrease
temperature by forced evaporation and than scan for Feshbach
resonances again to improve sensitivity.

\begin{figure}

{\centering \resizebox*{0.47\textwidth}{0.46\textheight}
{{\includegraphics{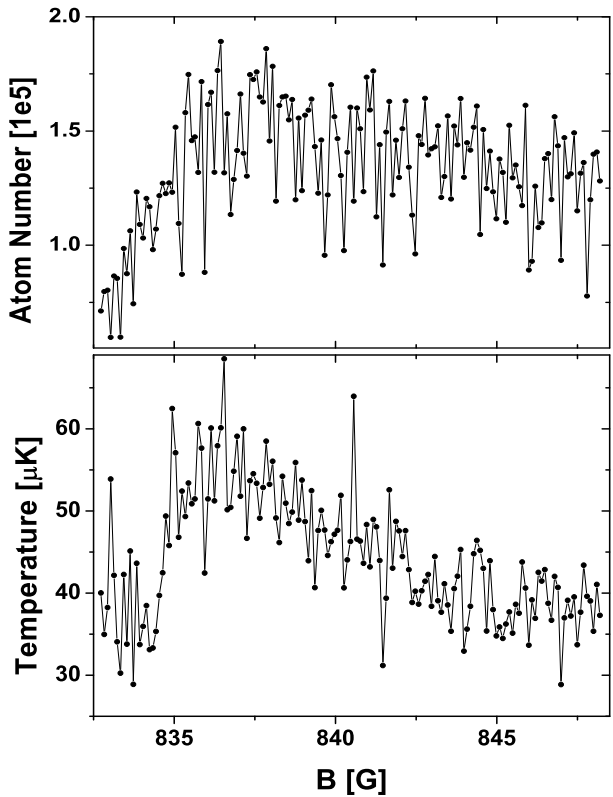}}}
\par}
\caption{\label{zeroX} The gain in atom number and increase in
temperature, which indicate the lack of free evaporation cooling,
reveal the zero crossing of the scattering length at $836\pm4$ G.}
\end{figure}

We execute a short forced evaporation of $1.5$ s reducing the trap
depth to $0.3$ mK, $15\%$ of its initial value. Evaporation is
performed with a bias field of $850$ G which is chosen to optimize
number-to-temperature ratio based on our scans for Feshbach
resonances at high temperature (Fig. \ref{feshbach}(b,c)). By the
end of this evaporation the atoms are cooled down to a temperature
of $\sim$50 $\mu$K. Our scan for inelastic losses in proximity to
Feshabch resonances reveals the same positions of the resonances
observed before including asymmetry in the profile of the broad
resonance. However, an additional feature has been identified which
we were unable to detect at high temperature. In Fig. \ref{zeroX},
zero crossing of the scattering length manifests itself in terms of
increased temperature and number of atoms \cite{O'Hara02}. The
hold-time at high bias field for this experiment is $0.5$ s. During
this time efficient free evaporation reduces the temperature to 40
$\mu$K if collisional cross-section is high enough. At zero crossing
the scattering length vanishes and so does the cross-section,
impeding efficient free evaporation. The asymmetry in this
measurement is caused by the proximity of the zero crossing point to
the narrow resonance at $831$ G. The zero-crossing point is detected
at $836\pm4$ G where the maximum temperature is observed.


\begin{figure}

{\centering \resizebox*{0.47\textwidth}{0.3\textheight}
{{\includegraphics{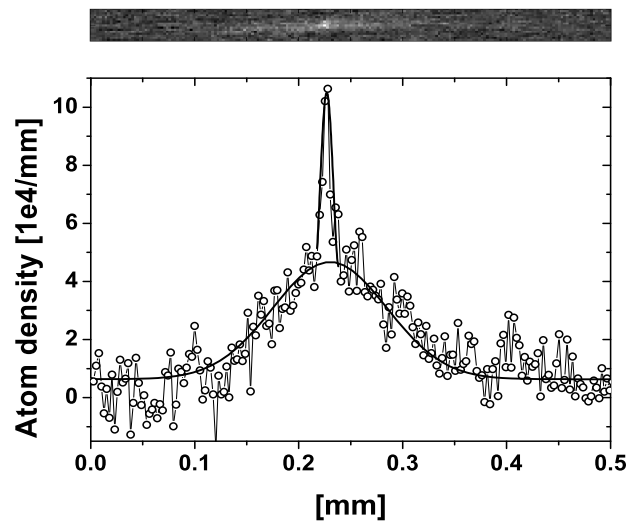}}}
\par}
\caption{\label{bec}On-set of BEC. \emph{In-situ} absorption imaging
of atoms at the BEC threshold. The thermal atomic cloud is fitted
with a Bose-Einstein distribution function which yields the
temperature of $380\pm40$ nK. The number of atoms in the BEC is
$\sim700$.}
\end{figure}

\section{Evaporation cooling to the BEC threshold}

Forced evaporation cooling of the optically trapped atoms down to
the BEC threshold is performed by attenuating the laser power, thus
reducing the trap depth which scales linearly with the power
reduction factor $\alpha$. A photo diode detector, located behind
one of the mirrors, collects a fraction of the trap beam's power
(see Fig. \ref{apparatus}). It generates a voltage readout that is
compared to a setpoint signal by a PID-controller. An error signal
is fed back to the RF-power supplier of the AOM. With this scheme we
are able to control trap depth reduction up to a factor of
$\alpha=3\times10^{-3}$. As is well known, trap oscillation
frequencies also decrease with the attenuation of laser power
($\propto\sqrt{\alpha}$), affecting re-thermalization efficiency
throughout the evaporation process. In addition to that, the strong
bias magnetic field, employed during evaporation, creates a weak
anti-trapping potential in the longitudinal direction. This
decreases further the optical confinement toward the end of the
evaporation. Indeed, in a single beam trap we were unable to cool
atoms below a temperature of 10 $\mu$K. The addition of the
zero-order diffraction beam solves this problem. At the beginning of
the evaporation it has no effect on the atoms as its potential depth
is negligible. However, with the reduction of the first order
diffraction intensity, it strengthen to create a confinement
potential to the atoms in the longitudinal direction of the trap.
The longitudinal oscillation frequency, determined by the confining
beam, is $2\pi\times60$ Hz. The zero order beam can be effectively
considered as a two dimensional confining potential because it
produces very weak ($0.14$ Hz) oscillation frequency in its
propagation direction which is easily overcome by the magnetic
anti-trapping potential. We note also that the crossed trap is
slightly shifted from the location of the zero order beam waist
which reduces somewhat the potential at the end of the evaporation.

Forced evaporation is found to work most effectively when the bias
field is set to $866$ G. Based on the theoretical curve (Fig.
\ref{feshbach}(a)) we estimate the scattering length to be
$(300\pm100)a_{0}$. The large uncertainty in the scattering length
is due to the uncertainty in the position of the broad resonance.
The trap depth is reduced exponentially in $3$ s with a time
constant of $330$ ms to less than $0.5 \%$ of its initial value. In
Fig. \ref{bec}, \emph{in-situ} absorption imaging of the trapped
atoms at $866$ G reveals the on-set of a BEC threshold by a familiar
bi-modal density distribution. The trace represents the atom
longitudinal density after integrating the radial direction of the
picture above it. Optical resolution is 4 $\mu$m, less than the size
of the BEC. The thermal atomic cloud is fitted with a Bose-Einstein
distribution function which reveals a temperature of $T=380\pm40$
nK. The total number of atoms is $6\times10^{3}$ which sets the
critical temperature to $T_{c}=350$ nK. The fitting of the BEC with
the inverted parabola of a Thomas-Fermi limit reveals $\sim700$
atoms in the condensate.

\section{Conclusions}

To conclude, we developed a method to all-optically condense
${^7}$Li atoms. This way facilitates the BEC production which is
extremely demanding in magnetic traps. We observed a BEC with
repulsive interactions on $|F=1,m_{F}=0\rangle$ state in less than
$3$ s of forced evaporation in a crossed beam optical dipole trap.
We use the tunability of the interatomic interactions in the
proximity of Feshbach resonances which we observed and characterized
for this internal state.

We note that the condensate life-time was very short, presumably
because of the extremely large scattering length. Moreover, our weak
optical trap at the end of the evaporation was not stable enough due
to the use of a single linear photo diode detector. A number of
improvements can be implemented to optimize the performance of the
described method. Better laser beam stability, which can be achieved
by using either two detectors \cite{Fuchs07} or a logarithmic
detector, and decrease of the scattering length toward the end of
evaporation will improve the condensate lifetime. Improved vacuum
would allow an increase in the number of atoms in the condensate.

We thank S. Kokkelmans for providing us with theoretical
calculations of Feshbach resonances on all sublevels of $^7$Li lower
hyperfine state. This work was supported, in a part, by the Israel
Science Foundation, through grant No. 1125/04.


\begin{thebibliography}{99}
\bibitem{Chapman01} M.D.~Barrett, J.A.~Sauer, and M.S.~Chapman,
Phys. Rev. Lett. {\bf 87}, 010404 (2001).

\bibitem{Yb_BEC} Y.~Takasu, K.~Maki, K.~Komori, T.~Takano, K.~Honda,
M.~Kumakura, T.~Yabuzaki, and Y.~Takahashi, Phys. Rev. Lett. {\bf
91}, 040404 (2003).

\bibitem{Cs_BEC} T.~Weber, J.~Herbig, M.~Mark, H.-C.~N\"{a}gerl, and
R.~Grimm, Science {\bf 299}, 232 (2003).

\bibitem{Granade02} S.R.~Granade, M.E.~Gehm, K.M.~O'Hara, and J.E.~Thomas,
Phys. Rev. Lett. {\bf 88}, 120405 (2002).

\bibitem{Kinoshita05} T. Kinoshita, T. Wenger and D.S. Weiss, Phys.
Rev. A {\bf 71}, 011602(R) (2005).

\bibitem{Dumke06} R. Dumke, M. Johanning, E. Gomez, J.D.
Weinstein, K.M. Jones, and P.D. Lett, New J. Phys. \textbf{8}, 64
(2006).

\bibitem{Fuchs07} J. Fuchs, G.J. Duffy, G. Veeravalli, P. Dyke, M.
Bartenstein, C.J. Vale, P. Hannaford, and W.J. Rowlands, J. Phys. B:
At. Mol. Opt. Phys. \textbf{40}, 4109 (2007).

\bibitem{Feshbach_review1} E.~Timmermans, P.~Tommasini, M.~Hussein,
and A.~Kerman, Phys. Reports {\bf 315}, 199 (1999).

\bibitem{Feshbach_review2} R.A.~Duine, and H.T.C.~Stoof, Phys. Reports {\bf 396}, 115
(2004).

\bibitem{Dalibard98} J.~S\"{o}ding, D.~Gu\'{e}ry-Odelin,
P.~Desbiolles, G.~Ferrari, and J.~Dalibard, Phys.~Rev.~Lett {\bf
80}, 1869 (1998).

\bibitem{Hulet95} C.C. Bradley, C.A. Sackett, J.J. Tollett, and R.G.
Hulet, Phys. Rev. Lett. {\bf 75}, 1687 (1995).

\bibitem{Li_a} E.R.I.~Abraham, W.I.~McAlexander, J.M.~Gerton, R.G.~Hulet,
R.~C\^{o}t\'{e}, and A.~Dalgarno, Phys. Rev. A {\bf 55}, R3299
(1997).

\bibitem{zero_a} J. Dalibard, in \textit{Bose-Einstein Condensation in Atomic
Gases}, Proceedings of the International School of Physics Enrico
Fermi, edited by M. Inguscio, S. Stringari, and C. Wieman (AIOS
Press, Amsterdam, 1999).

\bibitem{Schreck01} F. Schreck, G. Ferrari, K.L. Corwin, J.
Cubizolles, L. Khaykovich, M.-O. Mewes, and C. Salomon, Phys. Rev.
A {\bf 64}, 011402(R) (2001).

\bibitem{Wang07} R. Wang, M. Liu, F. Minardi, and M. Kasevich, Phys.
Rev. A \textbf{75}, 013610 (2007).

\bibitem{MOT98} U. Sch\"{u}nemann, H. Engler, M. Zielonkowski, M.
Weidem\"{u}ller, and R. Grimm, Opt. Commun. \textbf{158}, 263
(1998).

\bibitem{Mewes99} M.-O.~Mewes, G.~Ferrari, F.~Schreck, A.~Sinatra,
C. Salomon, Phys.~Rev.~A {\bf 61}, 011403(R) (1999).

\bibitem{Servaas} S.J.J.M.F. Kokkelmans, private communication.

\bibitem{bs_ENS} L.~Khaykovich, F.~Schreck, G.~Ferrari, T.~Bourdel,
J.~Cubizolles, L.D.~Carr, Y.~Castin, and C.~Salomon, Science
\textbf{296}, 1290 (2002).

\bibitem{bs_Rice} K.E.~Strecker, G.B.~Partridge, A.G.~Truscott and R.G.~Hulet,
Nature \textbf{417}, 150 (2002).

\bibitem{Fedichev96} P.O. Fedichev, M.W. Reynolds, and G.V.
Shlyapnikov, Phys. Rev. Lett. \textbf{77}, 2921 (1996).

\bibitem{FR_first_demonstration} S.~Inouye, K.B.~Davis,
M.R.~Andrews, J.~Stenger, H.-J.~Miesner, D.M.~Stamper-Kurn, and
W.~Ketterle, Nature {\bf 392}, 151 (1998).

\bibitem{D'Errico07} C. D'Errico, M. Zaccanti, M. Fattori, G.
Roati, M. Inguscio, G. Modugno, and A. Simoni, New J. Phys.
\textbf{9}, 223 (2007).

\bibitem{O'Hara02} K.M. O'Hara, S.L. Hemmer, S.R. Granade,
M.E. Gehm, J.E. Thomas, V. Venturi, E. Tiesinga, and C.J.
Williams, Phys. Rev. A \textbf{66}, 041401(R) (2002).




\end{thebibliography}
\end{document}